\newcommand{\boldsigma}{\mbox{\boldmath $\sigma$}}
\newcommand {\boldgamma}{\mbox{\boldmath$\gamma$}}

\newcommand{\bfx}{{\bf x}}
\newcommand{\bfy}{{\bf y}}

\def\poinc{Poincar\'{e} }
\def\bfq {{\bf q}}
\def\bfs {{\bf s}}
\def\bfn {{\bf n}}

\def\bfK{{\bf K}}

\def\bfk{{\bf k}}
\def\bfp{{\bf p}}  
 
\def\bfr{{\bf r}} 
\def\bfy{{\bf y}} 
\def\bfx{{\bf x}} 
\def\be{\begin{equation}}
 \def \ee{\end{equation}}
\def\bea{\begin{eqnarray}}
  \def\eea{\end{eqnarray}}

\newcommand{\bb}{\langle}
\newcommand{\kk}{\rangle}

\documentstyle[12pt,aps]{revtex}
\tighten
\def\poinc{Poincar\'{e} }
\def\bfq {{\bf q}}
\def\bfK{{\bf K}}

\def\bfk{{\bf k}}
\def\bfp{{\bf p}}  
\def\be{\begin{equation}}
 \def \ee{\end{equation}}
\def\bea{\begin{eqnarray}}
  \def\eea{\end{eqnarray}}

\title{\begin{flushright}{\normalsize NT@UW-03-07}\end{flushright}
Shapes of the Proton}
\author{Gerald A. Miller
\\ University of Washington
  Seattle, WA 98195-1560}

\begin{document}

\maketitle
\begin{abstract}
A model proton wave function, constructed using \poinc invariance, and
constrained by recent electromagnetic form factor data, is used to study
the shape of the proton. Spin-dependent quark densities are defined as
matrix elements of density operators in 
proton states of definite spin-polarization, and shown to have 
an infinite 
variety of non-spherical shapes. For high momentum quarks with spin parallel
to that of the proton, the shape resembles that of a peanut,
but  for quarks with anti-parallel spin the shape is that of a bagel.
\end{abstract}

\vskip0.5cm

Recent data \cite {Jones:1999rz,Gayou:2001qd} 
showing that  the ratio of the proton's electric and magnetic form
factor $G_E/G_M$, falls with increasing momentum transfer $Q^2$ for 
1$<Q^2<6 $ GeV$^2$
have created considerable attention. This behavior  or the equivalent    
statement 
 that  the ratio of Pauli
to Dirac form factors, $QF_2(Q^2)/F_1(Q^2)$ is a approximately constant,
indicates that quarks in the proton carry non-zero orbital angular momentum
\cite{ralston,Braun:2001tj,Miller:2002qb,Ji}.

This note is concerned with 
the shape of the proton. 
Several   subtleties and difficulties enter in determining the shape.
 As a particle of spin
1/2, the proton can have no quadrupole moment, 
according to the Wigner-Eckhart theorem.  Thus any non-spherical 
manifestation must reside in fluctuating components of the quantum wave
function. Another difficulty occurs in using experiments
in which momentum is transferred to the proton. The proton's final state  
 carries different total momentum than that of the initial
state, and the effects of boosts cause the initial and final state
wave functions to differ. For example, one often thinks  of a particle
in relativistic motion having a pancake shape because of the effects
of Lorentz contraction. Such an effect is caused by external influence
and is therefore not a manifestation of the intrinsic shape
of the proton. Another possible problem occurs through the use of 
light cone coordinates, which
  involves a separation of coordinates into
longitudinal and transverse that   complicates   a simple interpretation. Our
technique is to use a  model of the 
proton wave 
function\cite{Frank:1995pv},\cite{Miller:2002qb},\cite{Miller:2002ig}, 
constructed with \poinc invariance and  parameters
constrained by data, to compute
specific matrix elements using only the proton's rest  frame wave function 
and ordinary coordinates.

The basic  experimental observables  concerning us  are the
 form factors: 
\begin{eqnarray}
F_1(Q^2) ={1 \over 2P^+}\langle N,\uparrow\left| J^+\right| N,
\uparrow\rangle, \quad
Q\kappa  F_2(Q^2) ={-2M_N \over 2P^+}\langle N,\uparrow\left|
J^+\right| N,\downarrow\rangle,\label{defs}
\end{eqnarray}
with  $F_1$  obtained from a non-spin flip matrix element, and
$F_2$ from a spin-flip term. In evaluating the right-hand-side of
Eq.~(\ref{defs}) we take $J^+$ to be that of
free quarks\cite{cloud}, $\gamma^+$ times the quark charge. 
We use a frame in which 
$q^\nu=(p'-p)^\nu,\; Q^2=-q^2,\; q^+=0,\; \bfq=Q{\bf e}_x=-2\bfp$.

Our three-quark wave function is constructed using 
symmetries\cite{bere76,chun91}. It is
antisymmetric,  expressed in terms of relative
momentum variables, an eigenstate of the spin-operator defined
by the Pauli-Lubanski vector, is rotationally invariant and 
 reduces to non-relativistic SU(6) wave function in the limit that the
quark mass goes to infinity.
 The wave function is given by 
 \bea
\Psi_s(p_i)=\Phi(M_0^2)
u(p_1) u(p_2) u(p_3)\psi_s(p_1,p_2,p_3),\quad p_i=\bfp_i  s_i,\tau_i
\label{wave}\eea
where $\psi_s$ is a spin-isospin color amplitude factor\cite{chun91},
the $\bfp_i$ are expressed in terms of relative coordinates
(with  $-{\bf p}_{3\perp}={\bf K}$),  the
$u(p_i)$ are
  Dirac spinors and $\Phi$ is a spatial wave function. 
The arguments of the spatial wave function are
taken as the mass-squared operator for a non-interacting 
system\cite{Schlumpf:ce}:
\bea M_3\equiv 2(\bfk^2+m^2)^{1/2},\;
M_0=(\bfK^2+M_3^2)^{1/2}+ (\bfK^2+m^2)^{1/2},
\eea
where $\bfk$ is the relative momentum between quarks labeled 1 and 2
and $-\bfK$ is the momentum of the third quark\cite{explain}.
The spatial wave function is of the Schlumpf\cite{Schlumpf:ce} form: \bea
\Phi(M_0)={N\over (M^2_0+\beta^2)^{\gamma}}\;,  
\beta =0.607\;{\rm GeV}, \; \gamma=3.5,\; m = 0.267\; {\rm GeV}.
\label{params}\eea
 The value
of 
$\gamma$ is chosen that $ Q^4G_M(Q^2) $   is approximately  constant for
$Q^2>4\; {\rm GeV}^2.$
The values of $\beta$ and $m$ are determined by 
the    charge radius
and  magnetic moment of the proton. 
Slightly
 different values of $\beta,\gamma$ and $m$ 
are used when the effects of the pion cloud are incorporated\cite{Miller:2002ig}.
Using these newer values would 
 cause very small differences here.

The calculation of electromagnetic form factor  is
completely defined once 
 this  wave function and operator   $J^+$  are specified. Although the 
evaluation  was  presented 
long ago\cite{Frank:1995pv} and explained recently\cite{Miller:2002qb},
 it is worthwhile to briefly explain how the constant nature  of the
ratio $QF_2/ F_1$ emerges from the relativistic nature of the calculation.
 The wave function 
Eq.~(\ref{wave}) is completely anti-symmetric, so  
we may take $J^+$ to act only on the third quark 
which absorbs the momentum of the virtual photon. The average charge of
the third quark in 
the mixed-symmetric component of Eq.~(\ref{wave}) vanishes, so 
the only component of the
wave function that enters in the calculation of electromagnetic form factors
is the mixed anti-symmetric component in which the first two quarks 
have a vanishing total angular momentum.
 Then the spin of the proton $s$  is governed  by the third quark.
The relevant Dirac spinor is:
\bea
 u(p_3=\bfK,s)={1\over \sqrt{E(K)+m}}\left(\begin{array}{c}
(E(K)+m)\vert s\kk \\
\boldsigma\cdot\bfK\vert s \kk 
\end{array}
\right),\label{spinor}\eea
with $E(K)=(K^2+m^2)^{1/2}$.
The total angular momentum of the proton is denoted by $s$ and
the lower component contains a term  $\boldsigma\cdot\bfK$ that allows the
quark to have a spin opposite to that of the proton's total angular momentum.
The  vector  $\bfK$ reveals the presence of the quark orbital angular momentum:
the struck quark
may carry a spin that is opposite to that of the proton. Consequently  
helicity is not conserved\cite{ralston,Braun:2001tj,flip}.

Suppose $Q$ is very large compared to $\beta$ and
the third quark 
 changes its momentum from $\bfK$  to $\bfK'$. Then  
the 
form factor depends on the matrix element of $\bar{u}(K',s')\gamma^+u(K,s)$.
If we keep only the largest terms, those proportional
 to $Q$, we find
\bea
\bar{u}(K',s')
\gamma^+u(K,s)\sim\bb s'\vert Q +{K^+\over M_0}Qi\sigma_y \vert s\kk.\label{big}\eea
The spin-flip term proportional to $i\sigma_y$ arises from the lower component
of the spinor of Eq.~(\ref{spinor}), and this term has the same large factor $Q$
as the non-spin-flip term. Thus Eq.~(\ref{defs}) tells
 us that $QF_2$ and $F_1$ have 
the same dependence on $Q$, so      their ratio is     constant.
The construction of an eigenstate of spin, mandates the use of Dirac spinors.
The lower components of these spinors carry the quark orbital angular 
momentum responsible for the constant nature of $QF_2/F_1$.

 Our aim is to  
interpret these features of the wave function in terms of the shapes of the 
proton.  The technique of defining  spin-dependent density operators 
is  introduced by presenting
 two simple examples. Consider a non-relativistic nuclear wave function
consisting of a proton outside a $0^+$ inert 
core. The proton is bound by a combination of a central and 
spin-orbit potential, and  the single
particle wave function is an eigenfunction of total angular momentum.
Consider the case $(l,js)=(1,1/2s)$, with one unit of orbital angular momentum.
 Then the proton wave function is
a two-component Pauli spinor:
$\bb\bfr_p \vert\psi_{1,1/2s}\kk=R(r_p)\boldsigma\cdot\hat{\bfr}_p\vert s\kk,$
where $\bfr_p$ is the proton's position, and $\vert s \kk$ is
a  Pauli spinor. The charge density of this  system is the 
expectation value of the density operator $\delta(\bfr-\bfr_p)$:
with $
\rho(r)=\bb\psi_{1,1/2s}\vert\delta(\bfr-\bfr_p)\vert \psi_{1,1/2s}\kk=R^2(r),$
and is spherically symmetric.
However, consider instead the case that we require 
the proton at a position $\bfr$ to 
have a spin in a  direction defined by a unit vector ${\bfn}$.
 Then we define a spin-dependent  density:
\bea \rho(\bfr,{\bfn})=\bb
&&\psi_{1,1/2s}\vert\delta(\bfr-\bfr_p){(1+\boldsigma\cdot{\bfn})\over2}\vert \psi_{1,1/2s}\kk\nonumber\\
&&={R^2(r)\over 2}\bb  s\vert1+2\boldsigma\cdot \hat{\bfr}\;{\bfn} \cdot \hat{\bfr} -\boldsigma\cdot
{\bfn}\vert s\kk.\eea
An interesting special case is to take $\hat{\bfn}$
 parallel or anti-parallel to
the direction of the proton angular momentum 
$\bfs$. The direction of this vector defines an axis (the ``z-axis''), and
 the direction of vectors can be represented in terms of this axis:
$\hat{\bfs}\cdot\hat{\bfr}=\cos\theta$. With this notation
$ \rho(\bfr,{\bfn}=\hat{\bfs})={R^2(r)}\cos^2\theta,\;
 \rho(\bfr,{\bfn}=-\hat{\bfs}  )={R^2(r)}\sin^2\theta$ and
 the non-spherical shape is exhibited clearly. 
Note that the average of these
two cases  would give 
a spherical shape (as would an average over the direction of the total 
angular momentum),
 but the ability to
define a spin-dependent density allows 
the presence of the orbital angular momentum to be revealed in
 the shape of the computed density.

Another useful example is that of the 
Dirac electron wave function of the Hydrogen
atom. This wave function is a four-component spinor 
given by $\vert\psi_e\kk$ with
\bea \bb\bfr_e\vert\psi_e\kk=N\;r_e^{\gamma}\exp(-m_e\alpha\;r_e)\left(\begin{array}{c}
1\vert s\kk \\
i\alpha/2\boldsigma\cdot\hat{\bfr}_e\vert s \kk 
\end{array}\right),\eea
where $\alpha$ is the fine structure constant and $\gamma=\sqrt{1-\alpha^2}$.
We 
compute
the expectation value of the 
spin-dependent density operator, 
in terms of Dirac matrices:
$\delta(\bfr-\bfr_e))(1+\gamma^0\boldgamma\cdot{\bfn}\gamma_5)/2$, so that
with $\bfn=\hat{\bfs}, \;\rho(\bfr,\hat{\bfn}=\hat{\bfs})=
r^{-2\gamma}\exp{(-2m_e\alpha r_e)}\left[1+\alpha^2/4\cos^2\theta\right].$
We see that the Hydrogen atom is deformed! The angular dependence is
$\sim1+10^{-5}\cos^2\theta$ so  that the shape is almost spherical. But the
principle is clear: relativity as manifest by lower components of a Dirac
wave function  implies a deformed shape,
 if the matrix element 
is computed in a state of fixed total angular momentum.
For $\bfn=-\hat{\bfs},\;\rho(\bfr,\hat{\bfn}=-\hat{\bfs})=
\alpha^2\sin^2\theta$/4.
\begin{figure}
\unitlength1cm 
\begin{picture}(10,8)(0,-9)
  \includegraphics{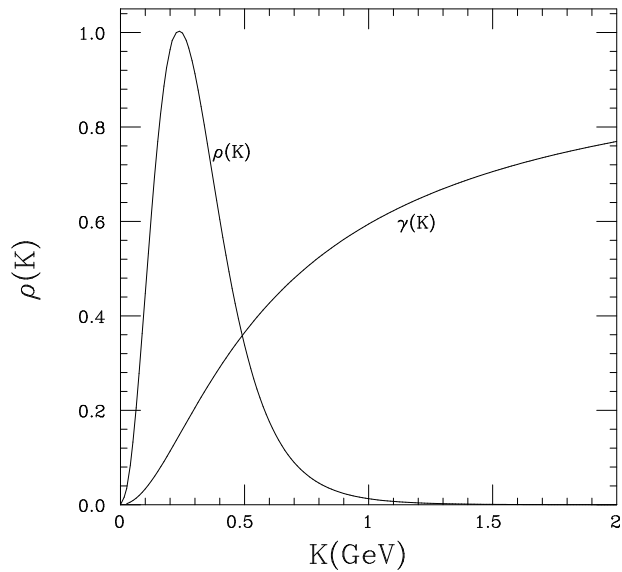}
\end{picture}
\label{fig:1}
\caption{Momentum probability $K^2\rho(K)$ in  dimensionless
units arbitrarily normalized.}
\end{figure}
With these examples in 
hand we turn to the proton. Its  wave function
 is specified in momentum space,  so  we define 
a momentum-space, spin-dependent,
  charge-density operator: 
\bea\hat{\rho}(\bfK,{\bfn})=\int {d^3r\over(2\pi)^3} e^{i\bfK\cdot\bfr}
\bar{\psi}(\bfr){\widehat{Q}\over e}
(\gamma^0+\boldgamma\cdot\bfn\gamma_5)\psi({\bf 0}), 
\label{qft}\eea 
where 
$\widehat{Q}/e$ is the quark charge operator in units of the proton charge.
The quark  field
operators are evaluated at equal time.
In QCD, one 
would need to add  a
factor between the 
field operators  to ensure gauge invariance.
With the present model, any gluons are contained within constituent
quarks, so the factor becomes unity. For our model, it is convenient to use 
first-quantized notation so that:
\bea\hat{\rho}(\bfK,{\bfn})
\equiv\sum_{i=1,3} (Q_i/e)\delta (\bfK -\bfK_i)
\left(1+(\gamma^0\boldgamma\cdot{\bfn}\gamma_5)_i\right)/2
.\label{rhohk}\eea We may compute probabilities for a quark to have
a momentum
$\bfK\equiv (K,\theta,\phi)$ and spin direction $\bfn$, for a spin-polarized
proton polarized in the $\hat{\bfs}$ direction. We find
\bea
\rho(\bfK,\bfn)=\bb\Psi_s\vert \hat{\rho}(\bfK,\hat{\bfn})\vert\Psi_{s}\kk
=\rho(K){1\over2}(1+\bfn\cdot\hat{\bfs}+{\gamma(K)}
(1-\bfn\cdot\hat{\bfs}+2\hat{\bfK}\cdot\bfn\hat{\bfK}\cdot\hat{\bfs}))
\label{shapek} \eea 
with \bea \rho(K)\equiv\int\;d^3k \Phi^2(k,K)(E(K)+m),\;\gamma(K)\equiv
{E(K)-m\over E(K)+m}.\label{rhodef}\eea
Some special cases of  Eq.~(\ref{shapek}) are interesting.
Suppose the  quark spin is parallel to the proton spin,  $\bfn =
\hat{\bfs}$,
then
$\hat{\rho}(K,\bfn =\hat{\bfs})=\rho(K)
\left(1+\gamma(K) \cos^2\theta\right)$. 
For small $K$ the shape is nearly spherical,
 but for large $K$ the $\cos^2\theta$ term becomes prominent. On the other
hand, the quark spin could be anti-parallel to the proton spin, 
 $\bfn =-\hat{\bfs}$. Then we find:
$\hat{\rho}(K,\bfn =-\hat{\bfs})=\rho(K)\gamma(K)
 \sin^2\theta,$ 
and the shape is that of a torus.
We may also take the
 quark spin perpendicular to the proton spin 
$\bfn\cdot\bfs=0$, so that
$
\rho(\bfK,\bfn\cdot\bfs=0)=\rho  (K) (1+\gamma(K))/2 +\gamma(K)
\sin\theta\cos\theta (\cos\phi n_x+\sin\phi n_y),$
to display the dependence on the azimuthal angle.
In each case, the non-spherical nature arises from the term proportional 
to $\gamma(K)$ caused by the lower components of the 
Dirac spinor (\ref{spinor}).

\vskip4.0cm
\begin{figure}
\unitlength1cm
\begin{picture}(10,8)(0,1)
\includegraphics{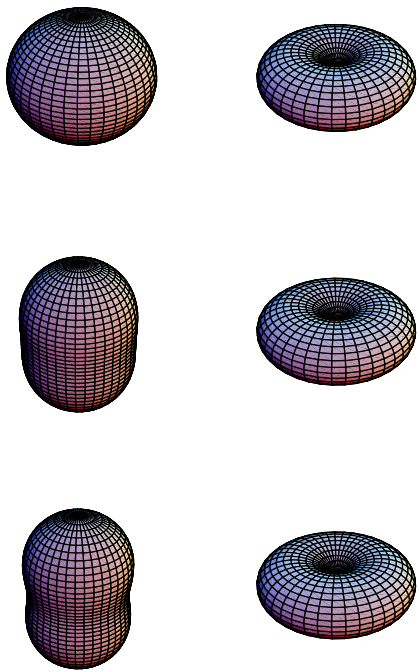}
\end{picture}
\label{fig:shapes}
\caption{Shapes of the proton. $\bfs$ is in the vertical direction $\bf\hat z$.
Left column quark spin parallel to nucleon
spin $\bfn=\hat\bfs$. Right column : quark spin anti-parallel to nucleon
spin $\bfn=-\bfs$. The value of $K$ increases from 0 to 1 to 4 GeV/c.  }
\end{figure}

We turn to the numerical evaluation and display of these shapes. 
The shape for a given value of $K$ is determined by the ratio
$\gamma(K)$ which reaches a value of 0.6 for $K=1$ GeV/c.
This implies considerable non-sphericity. 
The probability that a given value of $K$ is reached 
 is determined by the function
$K^2\rho(K)$, displayed in Fig.~1. The most likely value of $K$ corresponds
to $\gamma(K)=0.16$, and $\gamma(K)\le1$.

The shapes for the cases of quark spin parallel and anti-parallel to
the polarization direction of the  proton $\bfs$ are displayed in Fig.~2.
As the value of $K$ increases from 0 to 4 GeV/c the shape varies from
that of a sphere to that of a peanut, if $\bfn\parallel\bfs$. 
The torus or bagel shape is obtained if  $-\bfn\parallel\bfs$.
Taking $\bfn\perp\bfs$ leads to some very unusual shapes shown in Fig.~3.
\smallskip

\begin{figure}
\unitlength1cm
\begin{picture}(5,5)(-0,0)
\includegraphics{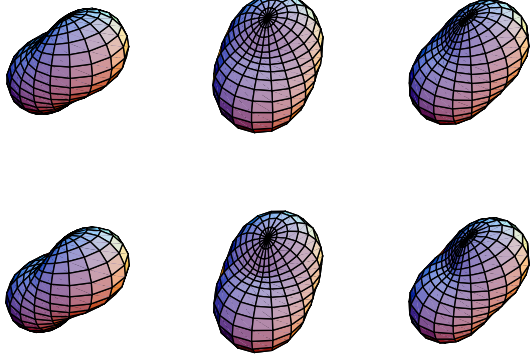}
\end{picture}
\caption{\label{fig:side}$\bfn\cdot \bfs=0$. 
Left column, $\bfn=\hat{\bfx}$ (out of page),
central: $\bfn=\hat{\bfy}$, right $\bfn=(\hat{\bfx}+\hat{\bfy})/\sqrt{2}  $.
 The momentum $K$ increases from
 1 to 4 GeV/c. }
\end{figure}

It is worthwhile to obtain a coordinate space look at the shape of the
proton. One can determine the probability for the separation of a quark 
from the center of mass to be 
$\bfr $  because  $\bfr$ is
canonically conjugate to $\bfK$.
We define 
a coordinate-space, spin-dependent,
  charge-density operator  that acts on the  quarks in the cm frame:
\bea\hat{\rho}(\bfr,{\bfn})
\equiv\sum_{i=1,3}
 (Q_i/e)\delta (\bfr -\bfr_i)(1+(\gamma^0\boldgamma\cdot{\bfn}{\gamma_5})_i)/2
\label{rhor}\eea so that we may compute probabilities for a quark to have
a position
$\bfr\equiv (r,\theta,\phi)$ and spin direction $\bfn$.
 We find
\bea
\rho(\bfr,\bfn)=\bb\Psi_s\vert \hat{\rho}(\bfr,{\bfn})\vert\Psi_{s}\kk
=\int\;d^3k\chi^\dagger(k,\bfr){1\over2}(1+\gamma^0\boldgamma\cdot{\bfn}\gamma_5)\chi(k,\bfr),
 \eea
 with
\bea &&\chi(k,\bfr)=\left(\begin{array}{c}
F_k(r)\vert s\kk \\
-i\boldsigma\cdot\hat{\bfr}G_k(r)\vert s \kk 
\end{array}\right),\eea
and $F_k(r)=\int d^3K\Phi(k,K)(E(K)+M)^{1/2}e^{i\bfK\cdot\bfr},
G_k(r)={\partial \over \partial r}\;\int d^3K\Phi(k,K){
e^{i\bfK\cdot\bfr}\over(E(K)+M)^{1/2}}.
$
We find
\bea
\rho(\bfr,\bfn)=\rho_U(r){1\over2}\left(1+\bfn\cdot\hat{\bfs}\right)
+\rho_L(r){1\over2}\left(1+2\hat{\bfr}
\cdot\hat{\bfs}\bfn\cdot\hat{\bfr}-\bfn\cdot\hat{\bfs}\right),
\label{shaper}\eea
where
$\rho_U(r)\equiv\int d^3kF_k^2(r),\; \rho_L(r)\equiv\int d^3kG_k^2(r)$. 
The pattern is similar to the one in momentum space, with
$\rho(\bfr,{\bfn}=\hat{\bfs})=\rho_U(r)+\rho_L(r)\cos^2\theta,\;\;
 \rho(\bfr,{\bfn}=-\hat{\bfs}  )=\rho_L(r)\sin^2\theta,
$ and $\rho(\bfr,\bfn={\hat\bfx+\hat{\bfy}\over\sqrt{2}} )={1\over2}\rho_U(r)+
{1\over2}\rho_L(r)\left(1+{2\over\sqrt{2}}\cos\theta\sin\theta(\cos\phi+\sin\phi)\right).
$

\begin{figure}
\unitlength1cm
\begin{picture}(10,8)(0,-8.5)
\includegraphics{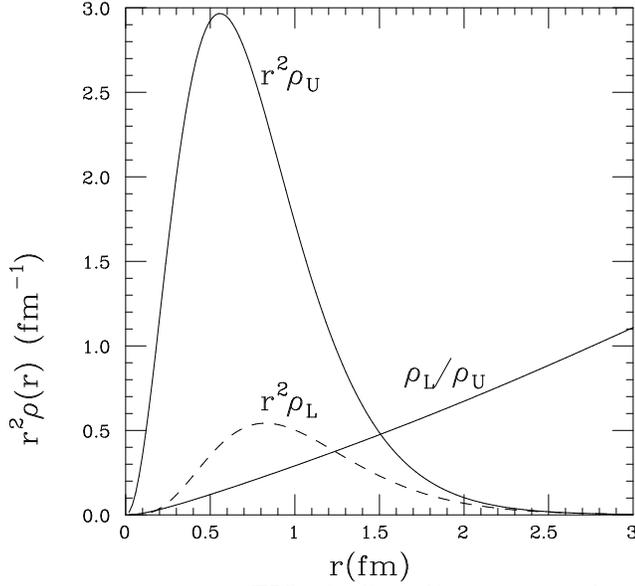}
\end{picture}
\label{fig:coord}
\caption{Coordinate space densities
}
\end{figure}

\noindent

The ratio $\rho_L/\rho_U$ controls the relativistic effects.
This can be much larger (Fig.~4) than the factor 
$\gamma(K)$ (limited by unity) controlling the momentum-space shapes, so
more extreme shapes are possible. The most likely value of $\rho_L/\rho_U$
is about 0.25, but there is no limit. 
The case with  $\rho_L/\rho_U$=3 is shown in 
in Fig.~5. A pretzel form is obtained if $\bfn$ is out of the page.
\vskip0.1cm

\begin{figure}
\unitlength1cm
\begin{picture}(3,3)(-1.0,1.)  
\includegraphics{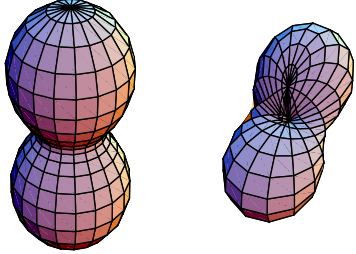}
\end{picture}
\label{fig:cshape}
\caption{Shape of the proton coordinate space. Left: $\bfn=\bfs$, right: $\bfn$
 points out of the page}
\end{figure}

The shape of the proton may be defined in terms of matrix elements of
spin-dependent density operators Eqs.~(\ref{rhohk},\ref{rhor})
 taken for protons in any fixed polarization
state. Relativity mandates the use of Dirac spinors to describe the 
quarks. These components, embodied in Eq.~(\ref{big}), lead to a
 constant ratio of
 $QF_2/F_1$ in accord with observation, and also to 
 shapes  of
  an infinite variety,
Eqs.~(\ref{shapek},\ref{shaper}), Figs.~2,3,5. 

We next consider  experiments  capable of  measuring  the 
 matrix element  $\hat{ \rho}(\bfK,\bfn)$,
Eq.~(\ref{qft}), for real
nucleons $\vert N\rangle$. Observe  that  $\int d^3K\hat{\rho}(\bfK,\bfn)$ 
 is a local operator. Its matrix element  is a linear combination of the
charge, integrals of spin-dependent structure functions $\Delta q$, and $g_A$,
it can be determined  from previous  measurements.
We find
\bea&& \int d^3K\langle N\vert \hat{\rho}(\bfK,\bfn=\pm \hat{\bfs})\vert N\rangle
={1\over 2}\langle N\vert \bar{\psi}(0){\hat{Q}\over e}(\gamma^0\pm\gamma^3\gamma^5)\psi(0)\vert N\rangle\nonumber\\
&&={1\over 2}\left(1\pm({1\over6}(\Delta u+\Delta d+\Delta s)+{1\over2}g_A\right)
=0.5\pm 0.34,\eea 
in which numerical values of $\Delta q$ are taken from Ref.~\cite{tw}.
The model we use gives $0.5\pm 0.37$ for the above quantity, indicating
that the results shown here may not be unrealistic.

The task of determining  $\rho(\bfK,\bfn)$  as a function of $\bfK$
 remains open.
However, the specific relativistic effects of orbital angular momentum
and the related      spin-flip effects
responsible for the non-spherical shapes can be expected to
influence  many measurable quantities.  These include:
spin-dependent structure functions $g_1,g_2$, the $N\Delta$ electromagnetic
transition form factor, and spin-dependent cross sections in high momentum
transfer scattering of  polarized protons. These topics will be addressed
in other publications.

\section*{Acknowledgments}
I thank the USDOE for partial support of this work. I thank M. Burkardt 
for useful discussions, and 
C. Glasshauser
and J. Ralston for emphasizing
 the importance of understanding the shape of the proton.


\begin{thebibliography}{9}
\bibitem{explain} 
Calculations of the form factors are simplified by using 
 light front dynamics to include the effects of
the boost. For our model the difference between the light front and
instant forms is contained in a coordinate transformation, which can be
made or not depending on the application.
 Here we will use $\bfq=0$,  so we  use the instant form.
\bibitem{flip} The ratio of $QF_2/F_1$ is a ratio of a 
 helicity-non-conserving  to a helicity conserving matrix element, if the 
reference frame of Eq.~(\ref{defs}) is used.
\end{thebibliography}
\end{document}